\documentclass{webofc}
\usepackage[varg]{txfonts} 
\newcommand{\be}{\begin{equation}}
\newcommand{\ee}{\end{equation}}
\newcommand{\bea}{\begin{eqnarray}}
\newcommand{\eea}{\end{eqnarray}}
\newcommand{\bwt}{\begin{widetext}}
\newcommand{\ewt}{\end{widetext}}
\newcommand\cC{{\cal C}}
\newcommand\cV{{\cal V}}

\newcommand{\nn}{\nonumber}
\begin{document}
\title{Holography, quantum complexity
and quantum chaos in different
models}
%
%

\author{\firstname{Dmitry} \lastname{Ageev}\inst{1}\fnsep\thanks{\email{ageev@mi.ras.ru}}}

\institute{Steklov Mathematical Institute of Russian Academy of Sciences}

\abstract{%
  This contribution to Quarks’2018 conference proceedings is based
on the talk presenting papers [1, 2] at the conference. These papers are devoted
to the holographic description of chaos and quantum complexity in the strongly
interacting systems out of equilibrium. In the first part of the talk we present
different holographic complexity proposals in out-of-equilibrium CFT following the local perturbation. The second part is devoted to the chaotic growth
of the local operator at a finite chemical potential. There are numerous results
stating that the chemical potential may lead to the chaos disappearance, and we
confirm the results from holography.

}
\maketitle
\section{Introduction}
\label{intro}

  The last decade AdS/CFT duality provided numerous connections of quantum information theory and gravity. The entanglement entropy, tensor networs, error correction codes have been described using this correspondence. The concept of entanglement has been shown to be described as the emergent phenomena from gravity degrees of freedom. This paper is devoted to the closely related topics in the recent studies of the AdS/CFT correspondence: the chaos and the quantum complexity. Recent works revealed different new measures of quantum chaos. Spectral form factors,  out-of-time-ordered correlators and operator size - all these measures are defined for quite general quantum theory \cite{Maldacena:2015waa}-\cite{Size1}. Also the concept of scrambling and scrambling time occurred to play an important role in the quantum system description. The scrambling time is defined as the time of the chaos onset and the scrambling is the phenomena of "informational smearing" over the system. 
  
  It occurs that the imprints of the chaotic  behaviour could be seen on the gravity side of the system. Moreover,  the holographic theories saturate the special bound on the chaotic measures of the system.  We investigate the effect of the chemical potential on the growth of the  operator size using the recent proposal by L.~Susskind \cite{Susskind1}-\cite{Susskind2}. This proposal relates the radial momentum of the bulk classical probe and the operator growth of the corresponding operator. The main result of  our observation tells, that chemical potential leads to the unbounded growth of the scrambling time of the charged observables, which indicates the chaos disruption above some critical parameters of the system. We list different chaotic models that show this behavour. These include the SYK with complex fermions \cite{Bhattacharya:2017vaz}, special melonic-dominated limit of the matrix quantum mechanics \cite{Azeyanagi} and particular random circuits \cite{Rakovszky:2017qit}-\cite{Pai:2018gyd}.  For 1/N structure of the SYK model see \cite{talk}.
  
  The notion of the complexity  in holography appeared after Susskind and Stanford \cite{Stanford:2014jda} noted some contradictory of different observables behaviour in the eternal black hole. While the most observable are scrambled at a logarithmic time the Einstein-Rosen bridge volume shows boundless and linear growth. This means that the system continue to evolve in some unusual way. More precisely it is formulated as: "the complexity of the system is still growing after the total information is scrambled". The description of the complexity in holography and quantum field theory attracted a lot of attention. The two most popular proposals in holographic definition of the complexity is the "complexity equals volume"(CV) and "complexity equals action"(CA) \cite{Brown:2015bva}. Both has advantages in its favor. It is important to study this proposal in the model which is well understood in the holographic and quantum field theory frameworks and has rich entanglement dynamics to compare complexity with it.    The important model of out-of-equilibrium conformal system is the 2D CFT following the local quench or in another words the CFT equilibrating after the local perturbation. We study the CA and the CV proposals for the total system and subregions. While  the CV conjecture lead to the behaviour similar to the entanglement  one can note some differences. The CA proposal for the subregions lead us to some reach pattern of behaviour qualitatively different from that the CV shows. Also it is worth to note that CA complexity saturates some conjectural bound on the rate of complexification at the initial moment of the local quench. This paper is organised as follows. In the Section 2 we describe the holographic setup of the local quench and calculate the complexity evolution in different proposals. In the Section 3 we describe how the chemical potential lead to the chaos suppression.
\section{The holographic complexity following the local quench}
\label{sec-1}

\subsection{Complexity in holography}
We start with the description of the main holographic complexity proposals.
The CV complexity  $\cC_{\cV}(t)$ is  proportional to the volume $\cV(B)$ of codimension-one bulk hypersurface $B$ attached to the fixed time $t$ slice of the boundary
 \bea
\cC_{\cV}(\Sigma)=\frac{\cV(B)}{G L},
 \eea
where $G$ is the gravitational constant, and $L$ is the characteristic scale of the bulk geometry defined below (for example the $AdS$ radius).

The CA duality relates complexity of the state to the value of certain gravitational action  evaluated on the Wheeler-DeWitt (WDW) patch
 \bea
\cC_{A}(t)=\frac{S({\cal W})}{\pi },
 \eea
where WDW patch $\cal W$ is the bulk domain of dependence of any Cauchy surface. Effectively this is the region bounded by null geodesics outgoing from the boundary at a time $t$.

\subsection{Holographic local quench and quantum complexity}

The holographic description of  the 2d CFT state evolving after the local perturbation is given  by the  $AdS_3$ Poincare patch deformed by the static massive particle  \cite{Nozaki:2013wia}. 
Mass $m$ of the  particle is then related to the conformal dimension $h$ of the perturbing operator.
The holographic metric describing  2D CFT local perturbation at time moment $t=0$ has the form
\bea \label{eq:long_metric}
&&ds^2=\frac{1}{z^2}\frac{\left(\alpha
   ^2 dx-2  t x dt+dx \left(u-z^2\right)+2 x z dz
   \right)^2}{\alpha ^4+2 \alpha ^2
   \left(u-z^2\right)+\left(z^2-v\right)^2}-\\\nn
&&-\frac{1}{z^2}\frac{\left(\alpha ^4+2
   \alpha ^2 \left(u+z^2(1-2 M) \right)+\left(z^2-v\right)^2\right)
   \left(\alpha ^2 dt+ \left(u+z^2\right)dt-2 t (xdx
   +zdz )\right)^2}{\left(\alpha ^4+2 \alpha ^2
   \left(u+z^2\right)+\left(z^2-v\right)^2\right)^2}\\\nn
&&\frac{1}{z^2}\frac{\left(\alpha ^4 dz+2 \alpha ^2 (udz -z(t dt +x dx ))+\left(v-z^2\right) \left(-2 t z dt +2 x z dx +\left(v+z^2\right)dz
   \right)\right)^2}{\left(\alpha ^4+2 \alpha ^2
   \left(u-z^2\right)+\left(z^2-v\right)^2\right) \left(\alpha ^4+2
   \alpha ^2 \left(-2 M
   z^2+u+z^2\right)+\left(z^2-v\right)^2\right)}
\eea
where $\alpha$ corresponds to the quench speed and $M$ is related to the particle mass $m$ as $M = 8GmL^2$.

It is known that the entanglement between the semi-infinite subsystems shows logarithmic late time growth. The complexity in our holographic setup is approximated by  the constant time slice volume form $\sqrt{\det \Sigma_t}$ of the metric is \eqref{eq:long_metric} at the fixed time $t$ . Let us define renormalized quantity $\Sigma=\Sigma(t,x,z,\alpha,M)$ as
\bea\label{DV}
\Sigma=\sqrt{\det \Sigma_t}-\frac{1}{z^2}.
\eea

Difference between the volume complexity and its equilibrium value is defined as $\Sigma$ integrated over the total time slice
\bea\label{CS}
\Delta{\cal C}(t)=\frac{2c}{3}\int\limits_{z>0}\int\limits_{x\in {R}} \Sigma dx dz,
\eea
where $c$ is the central charge of the CFT on the boundary.
 At early times ($t\rightarrow 0$)  the total systems complexity shows quadratic growth
\bea
\Delta {\cal C}(t)\approx 16 \pi h (1 +\frac{t^2}{2\alpha^2}).
\eea
 Note that the leading term in this expansion depends only on conformal dimension $h$. At  later times $\Delta {\cal C}$ must be computed numerically. In this section and in the next one, we use the entanglement entropy $\Delta S$ rescaled by the factor $1/4G$ and the complexity (calculated by formula \eqref{CS}) rescaled by $3/2c$.
\begin{figure}[t!]
\centering
\includegraphics[width=5.5cm]{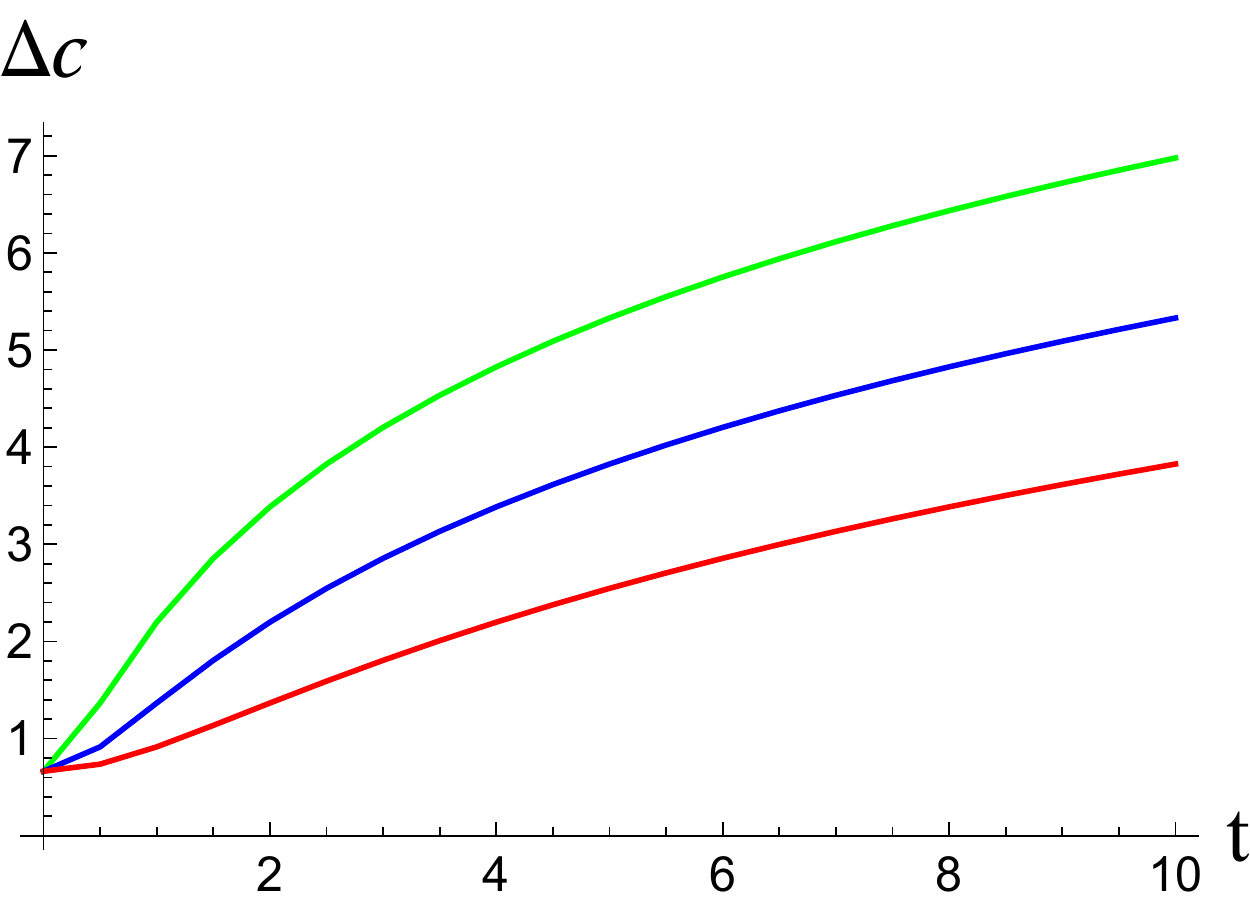}
\includegraphics[width=5.5cm]{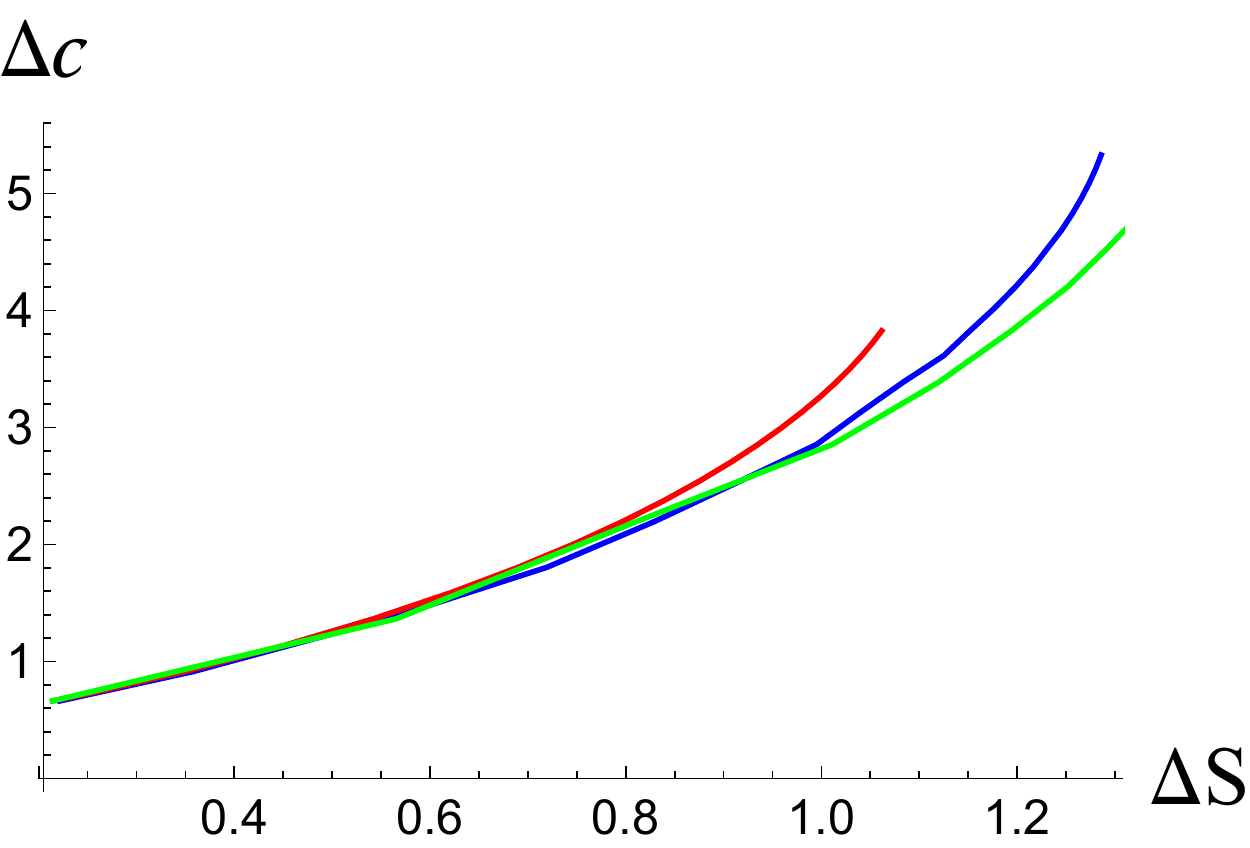}
 \caption{The left plot shows the time dependence of the (rescaled) complexity $\Delta \cC$ for the total system. The right plot shows the dependence $\Delta \cC(t)$ on (rescaled) $\Delta S(t)$ as time changes from $t=0$ to $t=10$. The green curves correspond to different  $\alpha=0.25$, the blue ones -- to $\alpha=0.5$, and the red one - to $\alpha=1$.}
 \label{fig:cwhole}
\end{figure}

In Fig.\ref{fig:cwhole} we plot the time dependence of $\Delta \cC $ for the total system, and compare it to the evolution of entanglement entropy of this bipartition. At the initial times the dependence of the complexity on the entropy is linear and universal for different $\alpha$.
Now we turn to the case of the CV complexity of the subsystem of the total system, namely the interval $x \in (-\ell,\ell)$.

\begin{figure}[t!]
\centering
\includegraphics[width=5.5cm]{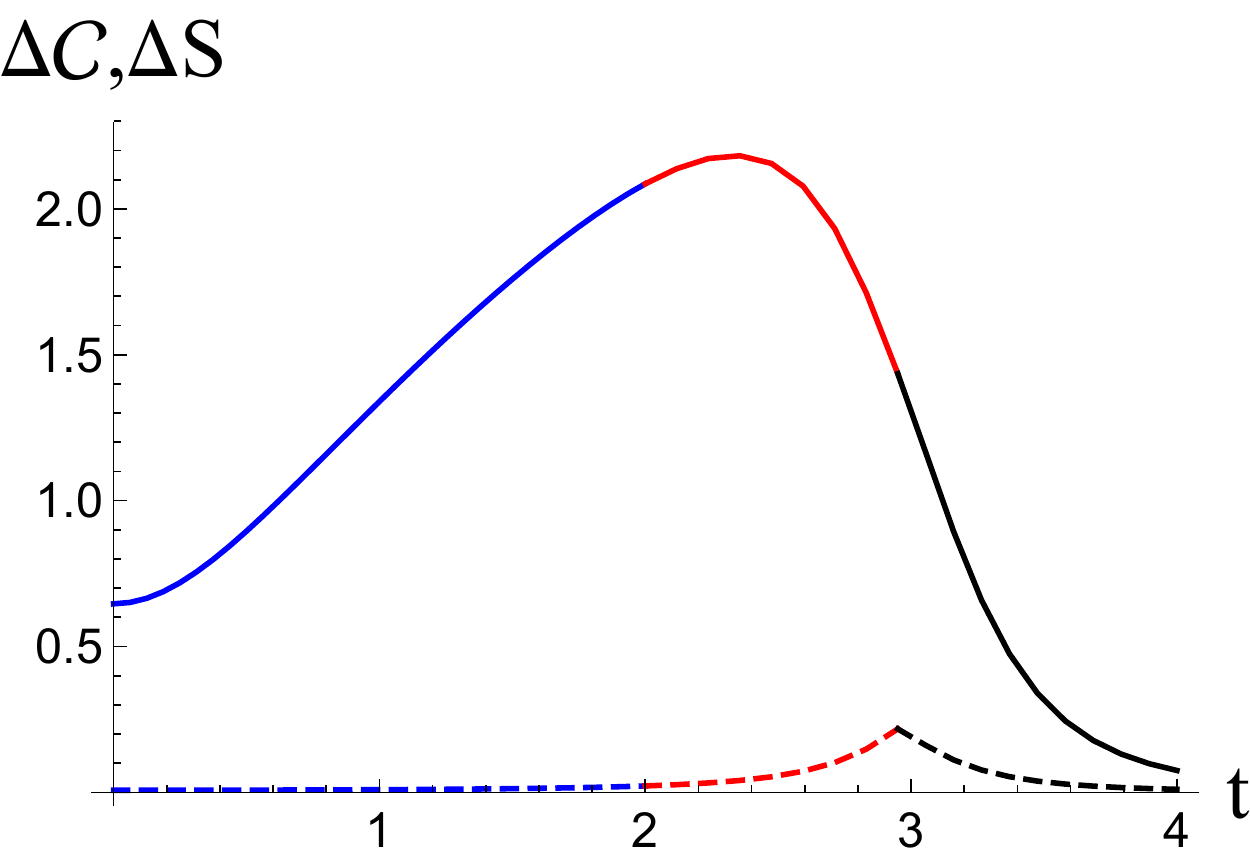}
\includegraphics[width=5.5cm]{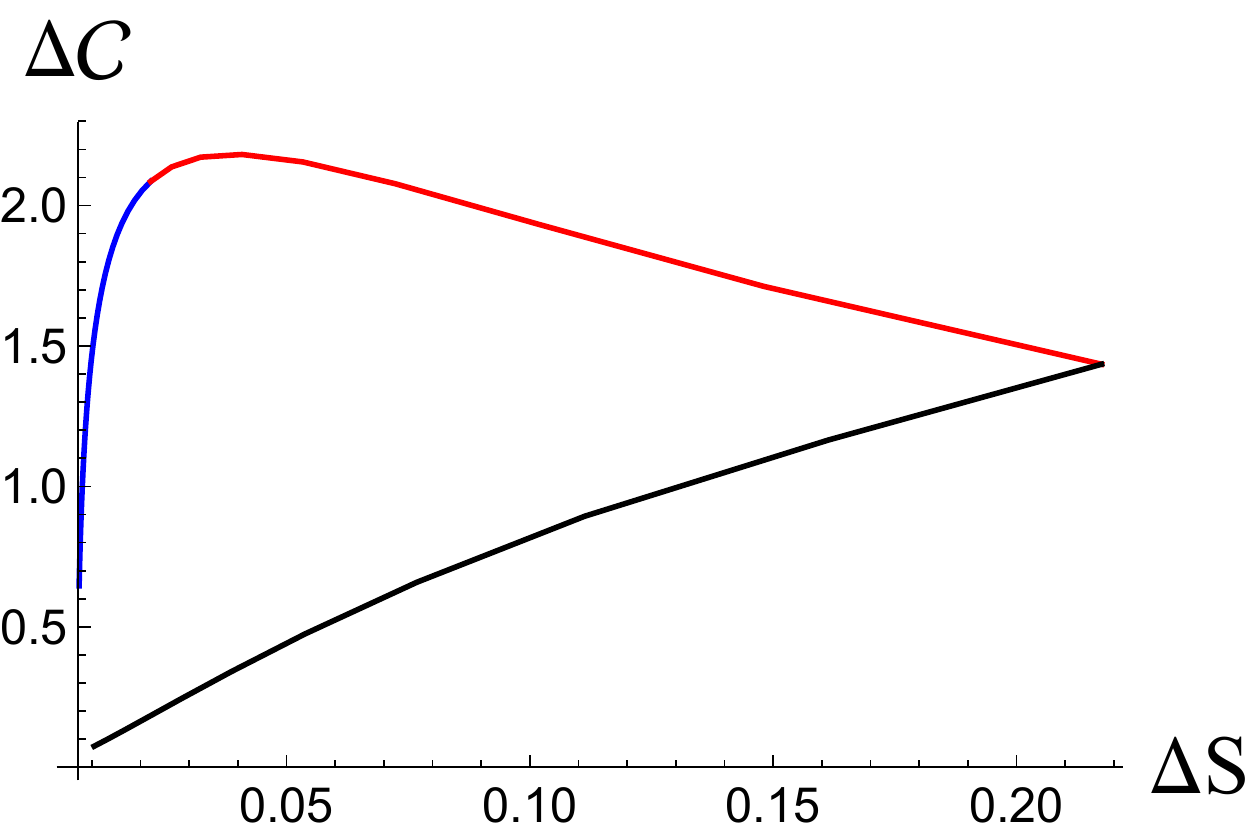}
 \caption{The left plot we present the time dependence of the (rescaled) complexity (solid curve) and the (rescaled) entanglement entropy (dashed curve) of $x\in(-3,3)$ interval for the same parameters. The right one is a parametric plot where we show the complexity $\Delta \cC$ of the quenched subsystem as a function of (evolving in time) entropy. The red, the blue and the black curves here correspond to the same time intervals as on the left plot.}
 \label{fig:CVell}
\end{figure}

The most interesting quench is the sharp one i.e. when $\alpha<\ell$. After the quench the complexity demonstrates quadratic growth, which is followed  by the regime of linear growth that continues  to the point where the complexity becomes maximal, see Fig.\ref{fig:CVell}. After passing the maximum, the complexity smoothly equilibrates.
We plot the complexity of the interval evolving after the quench in Fig.\ref{fig:CVell} and complexity dependence on the entropy. We see that in contrast to the entanglement of the same interval the complexity evolves smoothly, without sharp peak at the maximal value. Also complexity and entropy peaks occurs at different time moments. This lead to the different regimes of the dependence on the entropy.

Now we turn to the investigation of the CA complexity proposal  following the local quench. As stated above CA conjecture relates the complexity of the quantum  system state at time $t$ to the gravitational bulk action in a certain space-time region called Wheeler-DeWitt patch (WDW). This patch is the region bounded by light rays emanating from the boundary. 

\begin{figure}[h!]
\centering
a)\,\,\includegraphics[width=4.5cm]{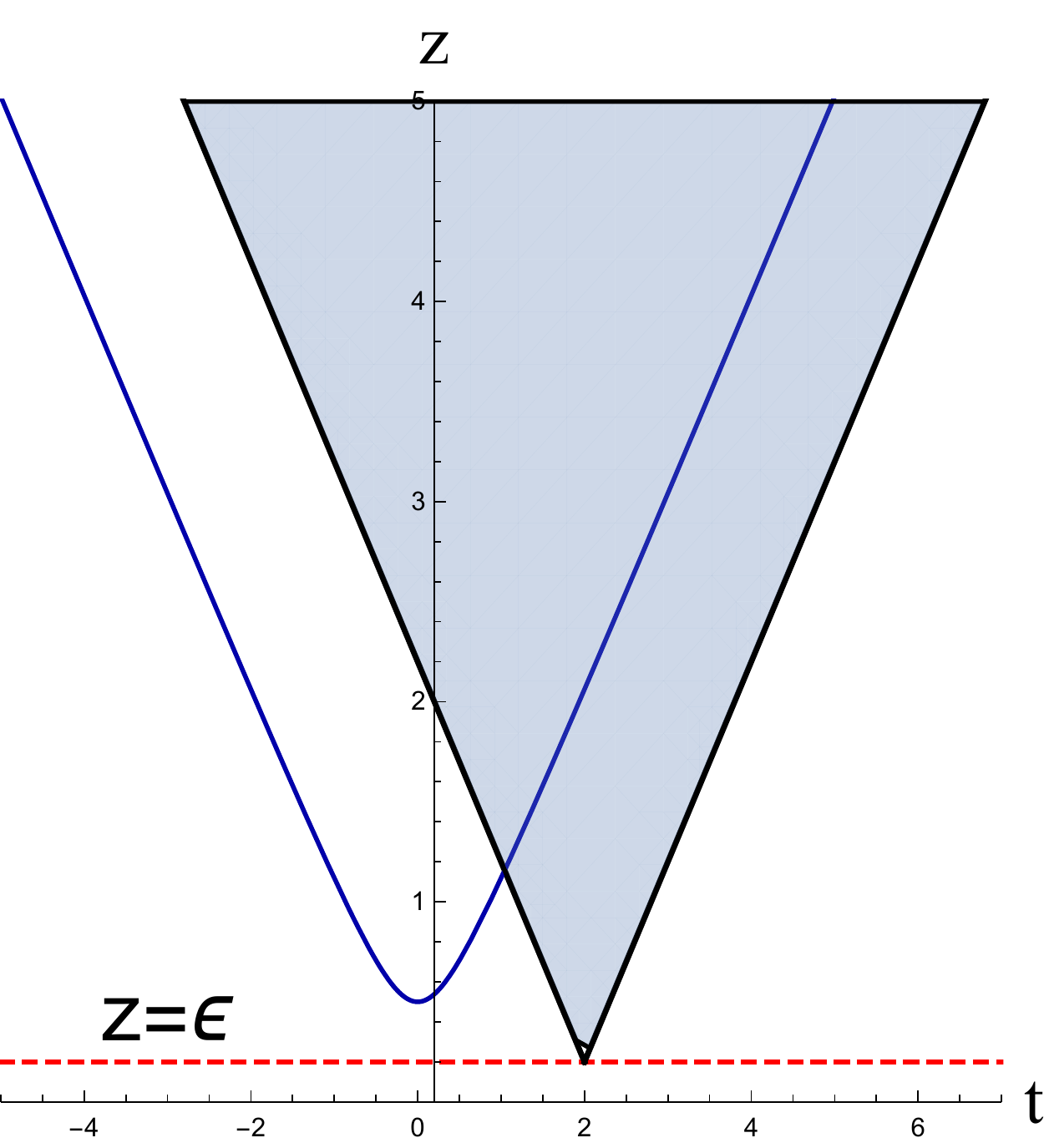}
b)\,\,\includegraphics[width=4.5cm]{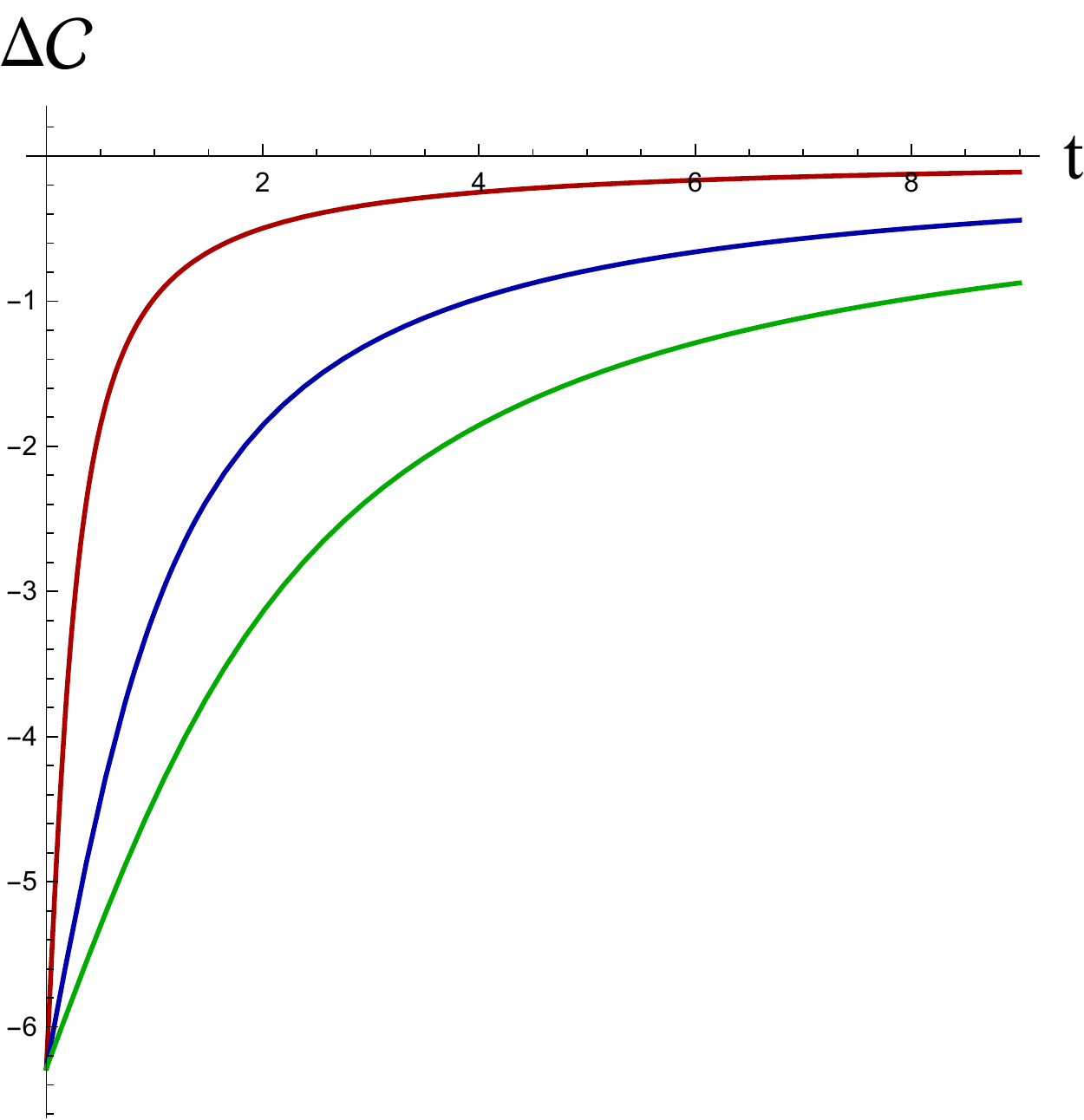}
 \caption{The left plot shows the WdW patch corresponding to the holographic local quench setup. The blue curve is the massive particle trajectory.  The red dashed line is UV regulator boundary, and the black solid lines are the boundaries of the WDW patch. The right plot shows the time evolution of the action complexity.  Here $\alpha=0.5,1,2$ from the bottom to the top respectively. One can see that instead of the unbounded growth (as in the CV case) the complexity returns to its unperturbed value.}
 \label{fig:WdW1}
\end{figure}

Working in the probe limit, we approximate the boundary of the patch to be the pure $AdS_3$ null rays. We take the regularization of the patch using  a small shift, $z_{bdy} = \varepsilon$, and consider null rays emanating from the boundary at time $t=\tau$: $z=\varepsilon+t-\tau$ and $z=\varepsilon-t-\tau$. We sketch the WDW patch in Fig.\ref{fig:WdW1}. The non-trivial time dependence in this approximation is due to the fact that at different times $\tau$ the part of particle trajectory intersects the WDW patch in a different way. 

The complexity  dependence on $\tau$ for $\tau>0$ has the form
\be\label{eq:CAtot}
\Delta\cC_{p}=-h  \left(1+\frac{2}{\pi} \arctan\left(\frac{\alpha }{2 \tau }-\frac{\tau }{2
   \alpha }\right) \right),
\ee
where we used the relation $m=2h$. Expanding this dependence when $t \rightarrow 0$, we get the linear growth of the complexity
\be \label{CAtotapp}
\Delta\cC_{p}\approx -2   h+ \frac{4 h  }{\alpha \pi }\tau=-2 h + \frac{2E}{\pi}\tau,
\ee
while at a late times we get asymptotic
\be 
\Delta\cC_{p}\approx -\frac{4 \alpha  h }{\pi \tau }
\ee
This calculation shows the intriguing result that can be stated as follows: the holographic complexity of the operator insertion quench saturates the Lloyds bound at the initial time moment.

\begin{figure}[h!]
\centering
a)\,\,\includegraphics[width=4.5cm]{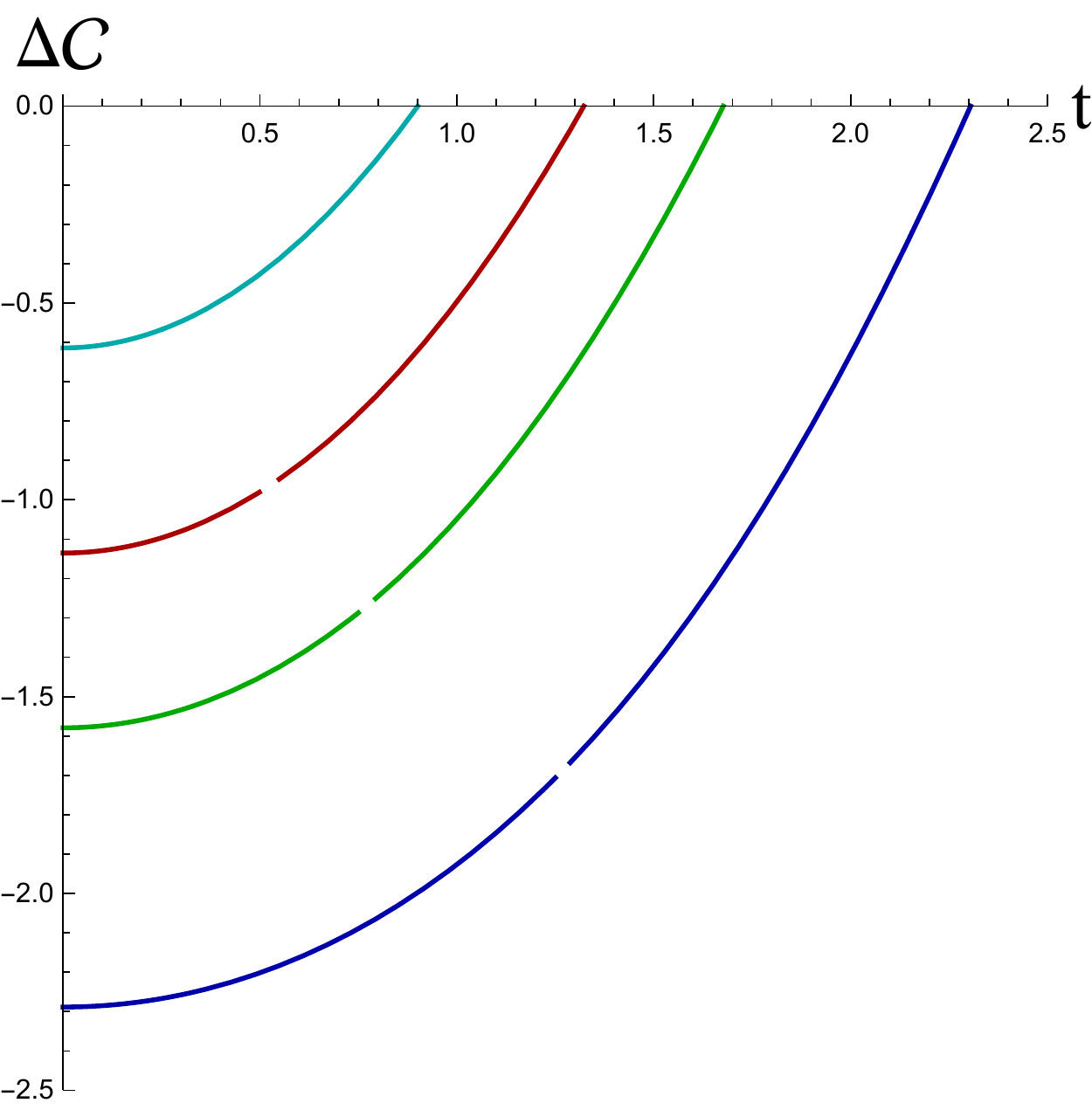}
b)\,\,\includegraphics[width=4.5cm]{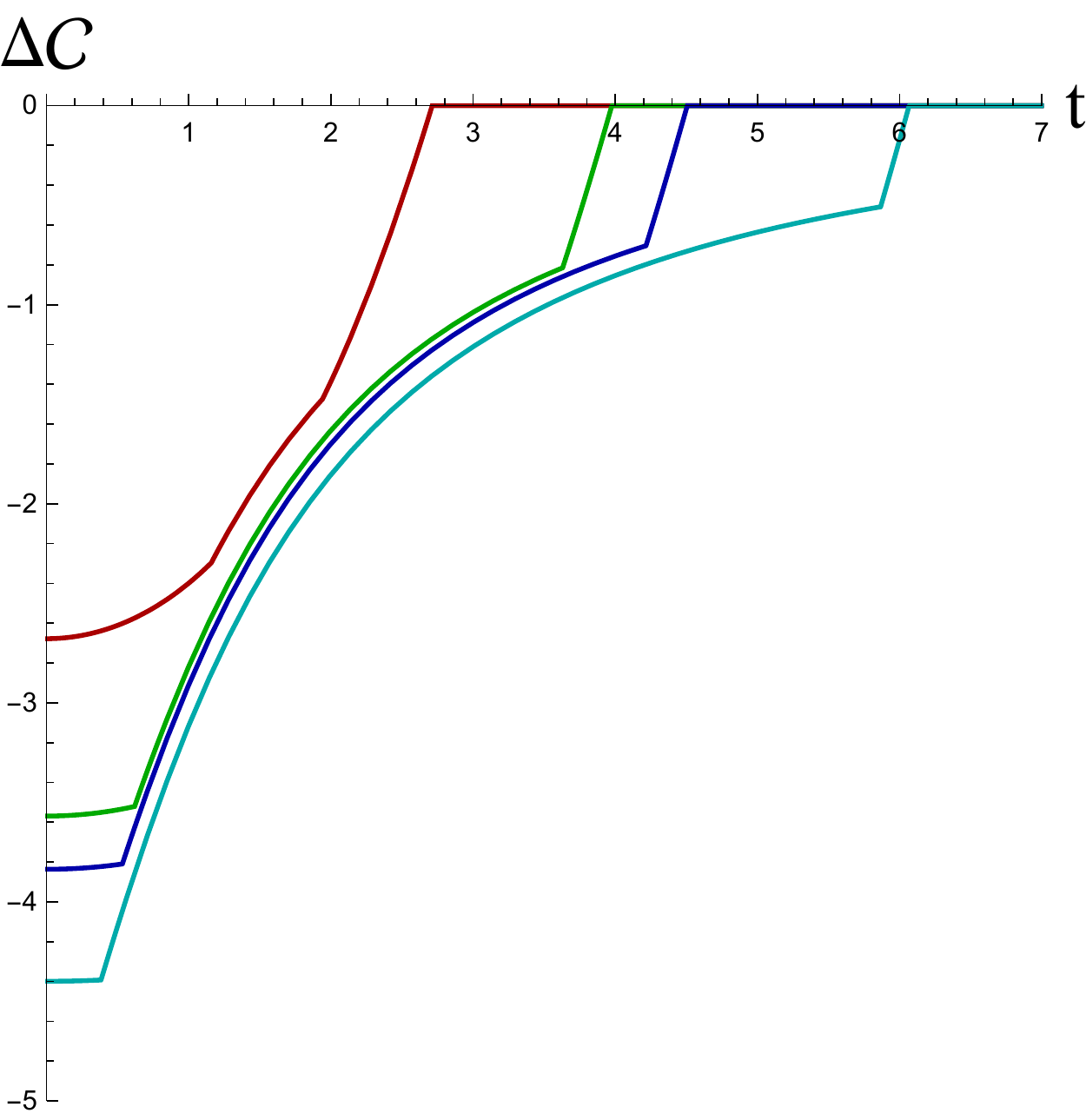}
 \caption{(a) Evolution of $\Delta \cC$ in the case of a mild quench, $\ell/4<\alpha<\ell/2$. Different curves correspond to $\ell=3.5,4,4.5,5.5$ (from top to bottom); $\alpha =1.5$. (b) The same quantity in the case of a sharp quench, $\alpha<\ell/4$. From top to bottom: $\ell=6.2, 8.5, 9.5, 12.5$; $\alpha=1.5$  }
 \label{fig:CAtl}
\end{figure}

Following this logic when calculating the CA complexity one can extend it to the case of the subregion complexity. The computation in this case is more complicated and i present only the result. Instead of the total WdW patch one has to calculate the action in the restricted region. The region  of interest is now the intersection of the WdW patch and the so called entanglement wedge. We present the complexity dynamics in Fig.\ref{fig:CAtl} for different intervals. We see, that the CA complexity "feels" some additional scale or parameter in the perturbed system that CV complexity or entanglement does not see.

\subsection{Finite chemical potential disrupts chaos}
\label{sec-2}

Now we turn to the second part of our talk that is devoted to the holographic description of the chaos. Our main interest here is the special quantity called operator size that is one of recently introduced chaotic measures. For the perturbation of a boundary theory by the time-evolving operator $W(t)$ and for the expansion it in the basis of some elementary operators $\psi_{a_s}$ as
\be 
W(t)=\sum_{s}W_s(t),\,\,\,\,\,\,W_s=\sum_{a_1<...<a_s} c_{a_1...a_s}(t)\psi_{a_1}...\psi_{a_s}
\ee
we define  $s$  as the operator size (see for example \cite{Size1,Shenker:2014cwa}).

\begin{figure}[h!]
\centering
\includegraphics[width=5cm]{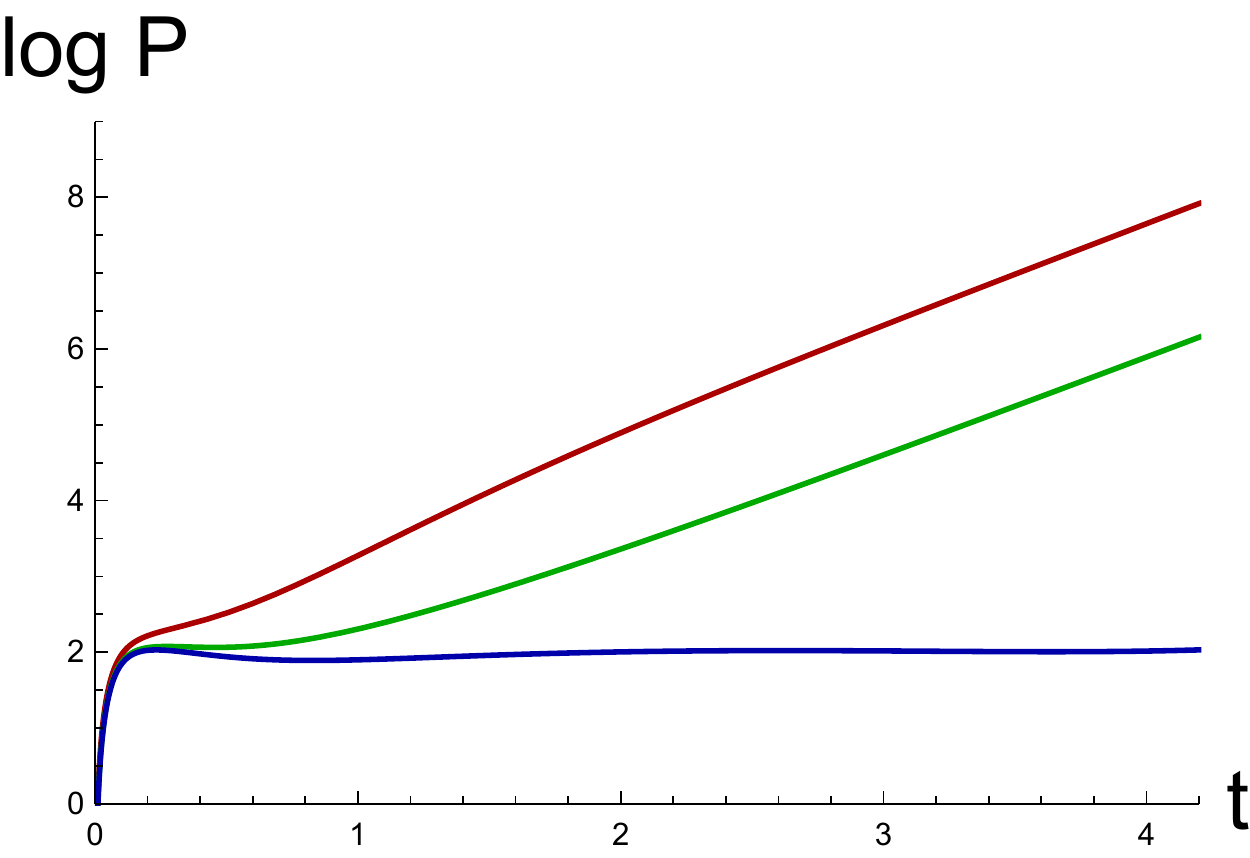}
\includegraphics[width=5cm]{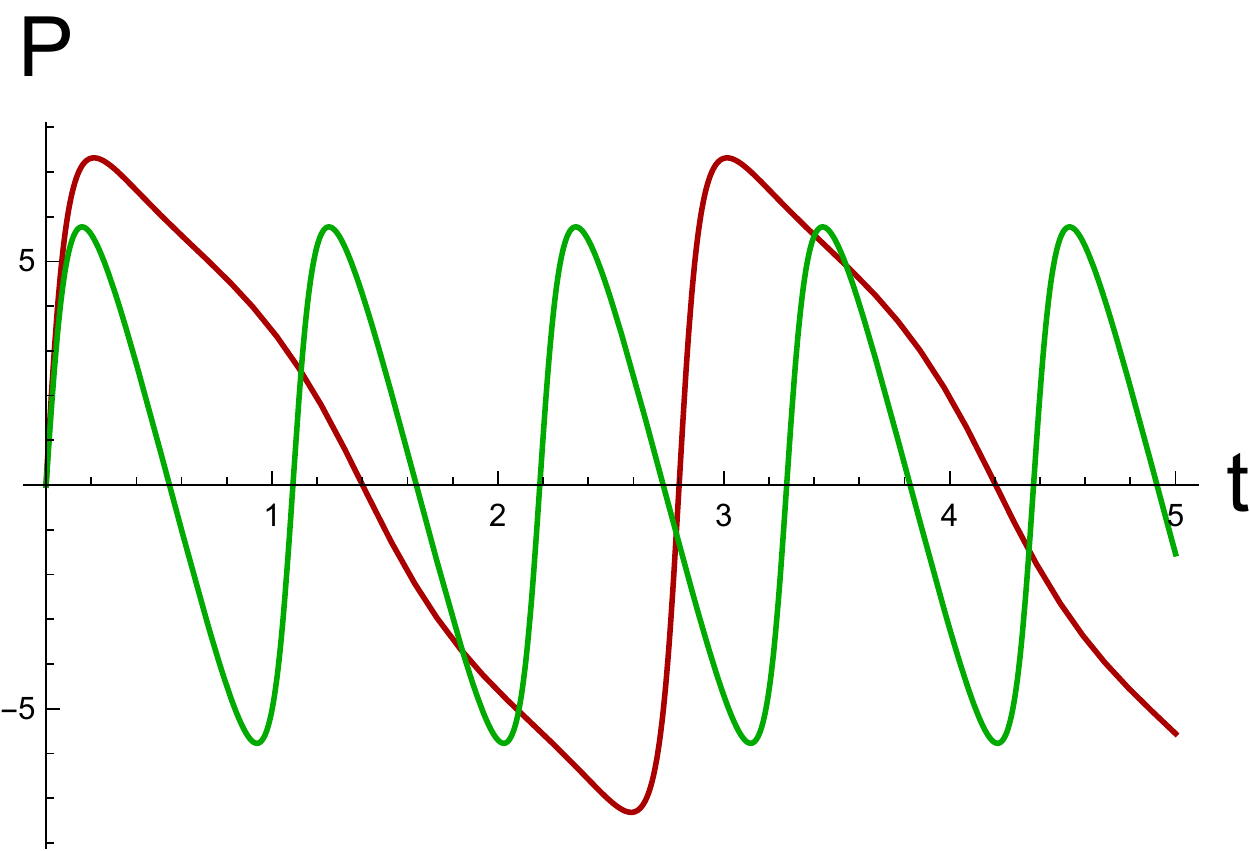}
 \caption{The momentum time dependence for different charge values and $\mu=1.3$, $d=3$, $z_h=1$, $z_*=0.1$. On the left plot red curve corresponds to the neutral particle, the green one to $q/q_{crit}=0.8$ and the blue one to $q=q_{crit}$.  On the right plot all curves correspond to the values above the critical charge. }
 \label{fig:nonextr-mom-low}
\end{figure}

In the chaotic theories the quantity is expected to have the exponential late time growth
\be s\sim\exp(\lambda_L t),
\ee
where $\lambda_L$ is called the Lyapunov exponent. It was shown, that $\lambda_L=2\pi T$ for holographic theories. Naively speaking this quantity shows how many "particles" have been "infected" by the initial perturbation. This quantity is quite "covariant" in the sense, that it involves only perturbing operator and the system state. 
Recently Susskind proposed the the relation between the particle falling in the black hole and the operator size. This correspondence could be formulated as
\be 
\text{\bf Operator size} \longleftrightarrow {\bf p_z(t)}.
\ee
The particle falling into the black solution will reach the Rindler region near the horizon and the particles momentum grows exponentially
\be 
p_z(t) \sim e^{\frac{2\pi}{\beta}t},
\ee
which is in accordance with the expectations from the QFT side.
We assume, that the charged particle that falls in the Reissner-Nordstrom-AdS black hole
 with the metirc and gauge field
  $A_t$ is given by \be\label{metr-2}
A_t=\mu\Big(1-\Big(\frac{z}{z_h}\Big)^{d-2}\Big),
\ee
with the metric  given by the Reissner-Nordstrom metric  
\be \label{metr-1}
ds^2=\frac{1}{z^2}\left(-f(z)dt^2+\frac{dz^2}{f(z)}+d\bar x^2 \right),\,\,\,\,
f(z)=1-M \Big(\frac{z}{z_h}\Big)^d+Q\Big(\frac{z}{z_h}\Big)^{2d-2},
\ee

 corresponds to the evolution of the charged observable.
The action of the charged particle has the  form
\be \label{Sq}
S=-m\int\sqrt{-g_{\mu\nu}\dot x^{\mu} \dot x^{\nu}}d\tau+q A_{\mu}\dot x^{\mu}d\tau,\,\,\,\,p_{\mu}=m \frac{g_{\mu\nu}\dot x^{\nu} }{\sqrt{-g_{\mu\nu}\dot x^{\mu}\dot x^{\nu}}}+q A_{\mu},
\ee
where $p_{mu}$ is the conjugated momentum.
For simplicity we take $m=1$ and $L=1$ without loss of generality.
We consider  the motion of the particle with the charge $q$ in this background.
For parametrization $z=z(t)$ we get the action
\be
S=-\int\frac{\sqrt{f(z)-\dot z^2/f(z)}}{z}dt+ q\mu\Big(1-\frac{z^{d-2}}{z_h^{d-2}}\Big)dt.
\ee
The energy $E$ of the particle is
\be
E=-\mu q\Big(1-    \frac{z^{d-2}}{z_h^{d-2}}\Big)+\frac{f(z)^{3/2}}{z(t)
   \sqrt{f(z)^2-\dot z^2}}.
\ee
 We are interested in the case of the positive particle charge  $q>0$.   If the energy is negative, the particle oscillates between two bulk points $z_{*,\pm}$. The critical charge when this takes place is given by the formula

\be
q_{crit}=\frac{\sqrt{f(z_{*,-})}}{ z_* A_t(z_{*,-})}.
\ee
The particle momentum is expressed as 
\be 
p_z=\frac{1}{z f(z)}\frac{\dot z}{\sqrt{f(z)-\dot z^2/f(z)}}.
\ee 
We plot the momentum time dependence for different values of the particle charge in Fig.\ref{fig:nonextr-mom-low}. From this figure we see, that above some charge the operator growth for the charged operator stops after some evolution. 
Finall let us briefly list the models from the quantum field theory that shows this effect of the chaos suppression

\begin{itemize}
\item The SYK model with complex fermion at a finite chemical potential
\item The special large D limit of matrix quantum mechanisc
\item The random circuits model with $U(1)$ conserving observables.
\end{itemize}

\section{Summary}
The local quench in 2d  CFT is the important example of the out-of-equilibrium system that has well-defined holographic dual. We studied how different complexity proposals work in the local quench setup. It would be interesting to study the complexity behaviour in the system with the bilocal quench \cite{AAT,AKT}. Let us list some notable features of the way how different complexity proposals work in the holographic local quench setup
\begin{itemize}
\item The CV complexity evolution is similar to the entanglement entropy, however it does not have sharp peaks. The initial growth is quadratic instead of the expected linear. The time when the complexity is maximal differs from the time when entanglement is maximal.
\item The CA complexity for the total system uniformly decrease. At the initial time the rate of complexification saturates the Lloyds bound. Also CA complexity is more qualitatively sensitive to the size of the subsystem.

\end{itemize}
In the second part of the paper we have shown how the suppression of the chaos by finite chemical potential can be seen from the holographic viewpoint. These results supports the recent Susskind's holographic proposal relating bulk probe radial momentum and the corresponding operator size.

\section{Acknowledgments}
I would like to thank the organisers of Quarks 2018 for possibility to give the talk and hospitality. The work is supported by the Russian Science Foundation
(project 17-71-20154).


\begin{thebibliography}{}

\bibitem{Ageev:2018msv} 
  D.~S.~Ageev and I.~Y.~Aref'eva,
  ``When things stop falling, chaos is suppressed,''
  arXiv:1806.05574 [hep-th].
  
\bibitem{Ageev:2018nye} 
  D.~Ageev, I.~Aref 'eva, A.~Bagrov and M.~I.~Katsnelson,
JHEP, 08 (2018), 071


\bibitem{Maldacena:2015waa} 
  J.~Maldacena, S.~H.~Shenker and D.~Stanford,
  JHEP {\bf 1608}, 106 (2016)

\bibitem{Shenker:2014cwa} 
  S.~H.~Shenker and D.~Stanford,
  JHEP {\bf 1505}, 132 (2015)

\bibitem{Size1} 
  D.~A.~Roberts, D.~Stanford and A.~Streicher,
  ``Operator growth in the SYK model,''
  arXiv:1802.02633 [hep-th].

\bibitem{Susskind1} 
  L.~Susskind,
  ``Why do Things Fall?,''
  arXiv:1802.01198 [hep-th].


\bibitem{Susskind2} 
  A.~Brown, H.~Gharibyan, A.~Streicher, L.~Susskind, L.~Thorlacius and Y.~Zhao,
  ``Falling Toward Charged Black Holes,''
  arXiv:1804.04156 [hep-th].


\bibitem{Bhattacharya:2017vaz} 
  R.~Bhattacharya, S.~Chakrabarti, D.~P.~Jatkar and A.~Kundu,
  JHEP {\bf 1711}, 180 (2017)
\bibitem{Azeyanagi} 
  T.~Azeyanagi, F.~Ferrari and F.~I.~Schaposnik Massolo,
  Phys.\ Rev.\ Lett.\  {\bf 120}, no. 6, 061602 (2018)

\bibitem{Rakovszky:2017qit} 
  T.~Rakovszky, F.~Pollmann and C.~W.~von Keyserlingk,
  ``Diffusive hydrodynamics of out-of-time-ordered correlators with charge conservation''
  arXiv:1710.09827 [cond-mat.stat-mech].
  
\bibitem{Khemani:2017nda} 
  V.~Khemani, A.~Vishwanath and D.~A.~Huse,
  ``Operator spreading and the emergence of dissipation in unitary dynamics with conservation laws,''
  arXiv:1710.09835 [cond-mat.stat-mech].

\bibitem{Pai:2018gyd} 
  S.~Pai, M.~Pretko and R.~M.~Nandkishore,
  ``Localization in fractonic random circuits,''
  arXiv:1807.09776 [cond-mat.stat-mech].

\bibitem{Stanford:2014jda} 
  D.~Stanford and L.~Susskind,
  Phys.\ Rev.\ D {\bf 90}, no. 12, 126007 (2014)

\bibitem{Brown:2015bva} 
  A.~R.~Brown, D.~A.~Roberts, L.~Susskind, B.~Swingle and Y.~Zhao,
  Phys.\ Rev.\ Lett.\  {\bf 116}, no. 19, 191301 (2016)

\bibitem{Nozaki:2013wia} 
  M.~Nozaki, T.~Numasawa and T.~Takayanagi,
  JHEP {\bf 1305}, 080 (2013)
\bibitem{AA}
 D.~S.~Ageev and I.~Y.~Aref'eva,
  Theor.\ Math.\ Phys.\  {\bf 189}, no. 3, 1742 (2016)
\bibitem{AKT}  I.~Y.~Aref'eva, M.~A.~Khramtsov and M.~D.~Tikhanovskaya,
  JHEP {\bf 1709}, 115 (2017)
  \bibitem{talk} I. Aref’eva, M. Khramtsov and M. Tikhanovskaya , “On 1/N diagrammatics in the SYK model beyond the conformal limit” Quarks 2018 proceedings
\end{thebibliography}
\end{document}